# INSPIRE: Evaluation of a Smart-Home System for Infotainment Management and Device Control

**Sebastian Möller[1], Jan Krebber[1], Alexander Raake[1], Paula Smeele[2], Martin Rajman[3], Mirek Melichar[3], Vincenzo Pallotta[3], Gianna Tsakou[4], Basilis Kladis[4], Anestis Vovos[4], Jettie Hoonhout[5], Dietmar Schuchardt[6], Nikos Fakotakis[7], Todor Ganchev[7], Ilyas Potamitis[7]**

[1] Institute of Communication Acoustics (IKA), Ruhr-University Bochum, Germany
[2] TNO Human Factors, Soesterberg, The Netherlands
[3] LIA, Ecole Polytechnique Fédérale de Lausanne (EPFL), Switzerland
[4] Knowledge SA (LogicDIS Group), Patras, Greece
[5] Philips Electronics Nederland B.V., Eindhoven, The Netherlands
[6] ABS Gesellschaft für Automatisierung, Bildverarbeitung und Software mbH, Jena, Germany
[7] Wire Communications Lab (WCL), University of Patras, Greece
sebastian.moeller@ruhr-uni-bochum.de

**Abstract**
This paper gives an overview of the assessment and evaluation methods which have been used to determine the quality of the INSPIRE smart home system. The system allows different home appliances to be controlled via speech, and consists of speech and speaker recognition, speech understanding, dialogue management, and speech output components. The performance of these components is first assessed individually, and then the entire system is evaluated in an interaction experiment with test users. Initial results of the assessment and evaluation are given, in particular with respect to the transmission channel impact on speech and speaker recognition, and the assessment of speech output for different system metaphors.

## 1. Introduction

Due to the increasing technological complexity of electronic devices in the home context, many users encounter severe problems in operating home appliances and services in their daily life. Intelligent interfaces which are easy to use may serve as a unique "home assistant" and hereby facilitate the operation (Wahlster et al., 2001). Such a home assistant is developed in the EU-funded IST project INSPIRE (INfotainment management with SPeech Interaction via REmote microphones and telephone interfaces; IST-2001-32746), making use of advanced speech technology. It consists of several components like acoustic pre-processing and adaptive noise cancellation, speech and speaker recognition, dialogue management, and speech output. With the help of the INSPIRE system, different devices in a living room can be controlled via speech, e.g. a TV, a video recorder, an answering machine, several lamps, a fan, and blinds. The system can be addressed in the home environment via a microphone array or a portable microphone, or from remote locations via the telephone network.

In order to assess the performance of the individual components and to evaluate their contribution to the quality of the entire system, five different test sites have been set up: Two sites where the signal pre-processing is analyzed under different acoustic conditions (WCL and ABS), a car-simulator site where the speech output component is tested (TNO), a home and office site where the interaction with the overall system is evaluated (IKA), and the Philips HomeLab (a Philips Research facility to test novel concepts in a real-home environment) where the final acceptability evaluation of the integrated system will take place. A detailed assessment and evaluation plan has been set up to co-ordinate the different actions. It consists of five steps:

1) Assessment of automatic speech recognition (ASR), speaker identification (SI) and speaker verification (SV) performance, taking the acoustic environment and the transmission channel into account.
2) Assessment and optimization of speech understanding performance.
3) Assessment of speech output quality for different system metaphors (see below), and in different acoustic environments and situations.
4) Evaluation of interaction quality in a simulated Wizard-of-Oz (WoZ) set-up.
5) Acceptability evaluation of the integrated system in the home scenario.

The final acceptability evaluation is still ongoing, and the optimization of the speech understanding module is very system-specific; thus, this paper will focus on the methodologies used in steps 1, 3 and 4 (Sections 3, 4 and 5), and it will present initial results obtained in the assessment and evaluation process. Prior to this, a brief overview of the INSPIRE dialogue system will be given.

## 2. System Overview

The INSPIRE prototype which has been set up so far can be accessed both from several points inside a house, and from remote locations through the telephone network or a packet-based network. It is available in two languages, German and Greek.

Inside the house, the user's speech is picked up via a wall-mounted microphone array or a wireless close-talking microphone. The room acoustic situation requires advanced beamforming, noise suppression, and echo cancellation techniques (Potamitis et al., 2003). Two commercial speech recognizers, one for Greek and one for German, are used. A purpose-built speaker verification and identification module prevents unauthorized use of the home system (Ganchev et al., 2002). When the system is accessed from remote locations, the transmission channel exercises a severe impact on speech recognition and speaker verification performance. In order to analyze this impact for a number of realistic transmission channels, a remote access simulation tool is connected to the INSPIRE system. It implements most of the degradations occurring in wireline, mobile and IP-based networks, including different user interfaces such as handset telephones, hands-free terminals, or headsets (Möller, 2000; Rehmann et al., 2002).

The dialogue with the user is managed via a distributed dialogue manager which co-ordinates a number of web services (Rajman et al., 2003). On the information output side, three different metaphors[1] have been implemented: A "talking head" metaphor which consists of a speaking avatar visualized on a wall-mounted screen, a "ghost" metaphor which consists of an immaterial assistant providing feedback through a set of loudspeakers in the whole home environment, and an "intelligent devices" metaphor where a specific voice is connected to each individual device and displayed locally (i.e., via one single loudspeaker). When calling from remote, only the "ghost" metaphor is used, because no position-related or visual information can be provided through the transmission channel. For each of the metaphors, either pre-recorded natural speech or synthesized speech can be selected. The speech is then displayed via loudspeakers mounted inside the house, or via the user interface connected to the remote access simulation tool.

## 3. Assessment of Speech and Speaker Recognition Performance

In order to assess the impact of room acoustics and transmission channels on speech and speaker recognition performance, a clean speech database has been collected. It consists of (a) 1370 utterances from prototypical dialogues recorded by 10 German speakers (5m, 5f), and (b) 500 dialogue utterances recorded by 10 Greek speakers (5m, 5f). The size of this database is relatively limited, because the assessment focuses on the influence of acoustic degradations, and not on the diversity of language and speakers. For the home environment, the database has been played back via a loudspeaker at different locations of a typical living room, and re-recorded via the microphone array in 35 conditions. The conditions differ with respect to speaker and microphone positions and to background noise. For remote access, the database has been played back by a head and torso simulator (HATS), and transmitted over the remote access simulation tool with 24 transmission channel settings, differing mainly with respect to the connected user interface, the background noise, the applied speech codec, and potential packet loss.

The performance of the Greek recognizer, expressed as the word accuracy (WA), is superior to the one of the German system for the clean speech data (98.7 vs. 69.0 % WA), and it shows a very high performance in most background noise conditions, see Table 1. The main reason for the superior overall performance of the Greek speech recognition module is the way the speech recognition engine is used: The Greek system uses a context-free grammar which anticipates the context for each function word, whereas the German recognizer is used in a key-word-spotting mode. A comparison experiment using the Greek recognizer without grammar (74.9% WA) shows that the context accounts for almost 24% WA difference; the rest will be due to the speech recognition module itself, and perhaps to language characteristics.

---

[1] The term "metaphor" is used here to describe the transfer of meaning to the machine interaction partner by the human interaction partner, due to the similarity to a human partner in its apparent shape, in its function, and in its use.

| Background noise | Speaker position relative to the microphone array | |
|---|---|---|
| | 0° | 45° |
| No noise | 93.4 % | 97.3 % |
| Music noise, 6 dB SNR | 94.4 % | 96.2 % |
| Speech babble, 6 dB SNR | 90.3 % | 86.0 % |

Table 1: Word accuracy for the Greek recognizer under different background noise conditions. Speaker at 2.7 m (0°) and 3.0 m (45°) distance from the microphone array, background noise source at 3.0 m and 55°.

The effect of transmission channel degradations, including the acoustic effects of the connected user interfaces, is clearly observable for both the German and the Greek recognizer. For a default transmission channel setting with a handset telephone, a 300-3400 Hz bandpass limitation, and a low circuit noise floor, WA is degraded by approx. 1% for the Greek recognizer, and by almost 14% for the German recognizer. As it was expected, the handset and the headset are the best-performing user interfaces under all background noise conditions. The performance for the hands-free terminal depends on the recognizer and on the acoustic conditions in the room (terminal set-up and background noise).

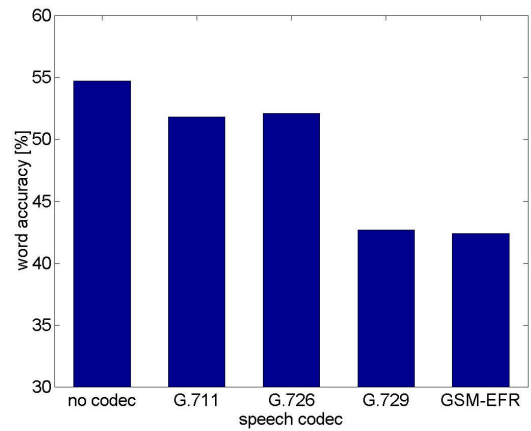

Figure 1: Word accuracy of the German recognizer for different speech codecs

For speech codecs, both recognizers seem to cope well with the G.711 and G.726 coding algorithms, see Figure 1. The G.729 and the GSM-EFR codecs result in a stronger degradation. For G.711 and G.729, packet loss leads to a decrease in recognition performance which is almost linearly related to the percentage of lost packets. These results are not depicted here to save space.

The speaker recognition experiments confirmed that the SV (remote access) and SI modules (home environment) perform well. Both modules had a relatively high performance in the clean speech scenarios: 98.2% (Greek) and 93.3% (German) for the SV, and 98.7% and 94.3% for the SI, respectively. In remote access scenario, SV performance is reduced by about 3-5%, depending on the degradations introduced by the transmission channel and the characteristics of the user interface. In the home environment, SI performance was reduced by room reverberation (a decrease of about 4-6%), variations in the speaker's position (decrease by about 2%), and the presence of concurrent speakers or background noise. It was found that music has a more severe effect on the SI performance (decrease of 2-18%) than the concurrent speaker (1-6%), and that the decrease depends on the actual SNR, the noise source position, and the language. For instance, for the Greek language, background music noise with 6 dB SNR reduces the SI scores to 85-87%, and a concurrent speaker with the same SNR to 89%.

## 4. Assessment of Speech Output Quality

The speech output quality of the overall system can only be assessed in a realistic application scenario. For each of the metaphors described above ("ghost", "talking head" and "intelligent devices"), a number of typical sentences have been recorded, both from 5 natural speakers (3 female and 2 male) and from a commercial TTS system (male voice). These stimuli were played back to the test subjects in three different environments: (1) A living room, implementing the characteristics of the three metaphors (directed or diffuse loudspeaker representation, partly with a talking-head display) and a feedback from the devices addressed by the stimuli (e.g. the blinds of the test room are manipulated when a message regarding the blinds is played back); (2) an office environment where the speech files were first transmitted through the remote access simulation tool and then presented via different user interfaces (handset, headset and hands-free terminal); and (3) a car simulator where the test subjects had to perform a driving task while listening to the stimuli through the car loudspeakers. In (2) and (3), only the "ghost" metaphor could be tested.

After listening to a speech sample, the test subjects were first asked about the device type and function which were mentioned in the speech stimulus. In this way, subjects were forced to concentrate on the content of the speech signal, and not on the surface form alone. Secondly, they had to rate their overall impression, the required listening-effort, the voice pleasantness, and the adequacy of the voice for the described service. Continuous rating scales which were labeled with five attributes have been used for this purpose, similar to the ones given in ITU-T Rec. P.851 (2003). Mean judgements and standard deviations have been calculated from this data, the value 5 corresponding to the most positive and 1 to the most negative attribute.

26 subjects participated in the living-room experiment. They were between 20 to 76 years old (mean: 34.0 years). In the office environment, 24 subjects took part. Their age ranged between 20 and 70 years, with a mean of 31.6 years. About half of the test subjects in the living-room and office environments can be regarded as having some prior knowledge of speech technology. In the car environment, only 16 out of 21 test subjects successfully completed the experiment (5 suffered from motion sickness). 6 of these subjects were male and 10 female. They were between 21 and 64 years old (mean: 32.7 years).

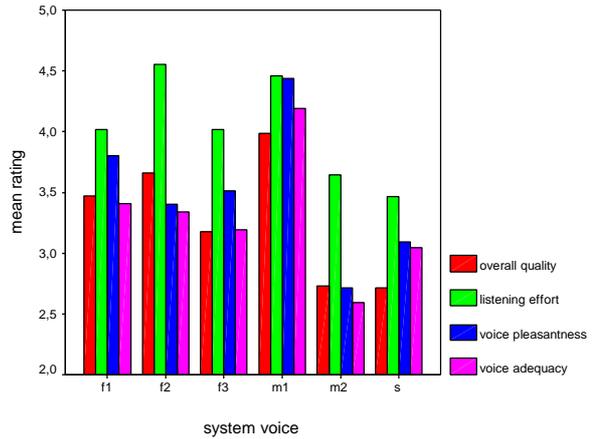

Figure 2: Mean ratings for different voice options in the living-room environment, "intelligent devices" metaphor

Figure 2 shows the ratings for the different naturally produced (*f1*, *f2*, *f3*, *m1*, *m2*) and the synthesized (*s*) system voices, averaged over conditions with and without background noise in the living room. Although the overall quality of the synthesized voice is amongst the lowest observed in the tests, there is no particular gap to the quality level reached by naturally produced voices. In particular, voice pleasantness and adequacy are rated higher for the *s* than for the *m2* voice. The low overall quality rating of the synthesized voice mainly seems to be due to the required listening-effort.

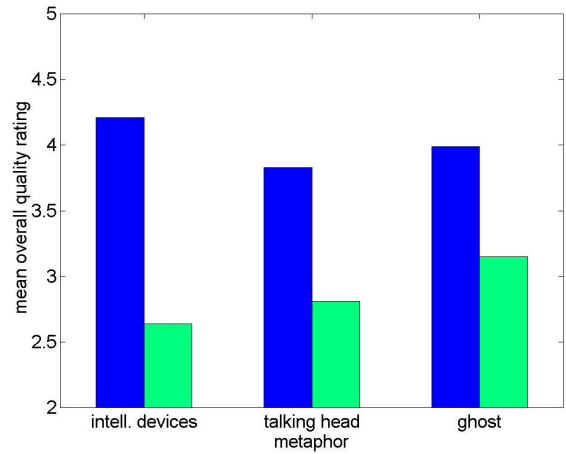

Figure 3: Comparison of overall quality ratings across metaphors for the *m1* (left bars) and *s* voices (right bars).

The natural voice *m1* and the synthesized voice *s* have been used in all three metaphors. A comparison between the mean overall quality ratings obtained for these voices shows that the synthesized voice is rated worse in the "intelligent devices" metaphor (compared to the "ghost" and the "talking head"), whereas the opposite is the case for *m1*. Apparently, the connotation with different devices leads to more distinguished ratings for the individual voices compared to a single (visible or invisible) assistant. For a single home assistant, the ratings for the individual voices are closer together on the scale. The visual information linked to the talking head does not seem to provide any advantage in the present case (e.g. due to a higher intelligibility). It is assumed that the lack of synchronicity between the lip movements and the acoustic signal is responsible for this finding.

Comparing the three different test environments, the effects of the transmission channel in case of remote access could clearly be observed both on the synthesized and on the naturally produced speech. The driving task did not significantly influence the quality ratings. The cognitive load of this task (driving on a moderately frequented suburban road with various events) might have been below a critical threshold for all voices considered in the test.

## 5. Overall System Evaluation

The overall system can only be evaluated in a realistic environmental setting, with test subjects performing actual tasks. These tasks have been defined beforehand and are conveyed with the help of scenarios, i.e., short stories of envisioned real-life situations. Scenarios illustrate the usage, the functionality and the purpose of the addressed devices. Test users followed three of these scenarios, and judge different aspects of perceived quality on a specifically designed questionnaire. The questionnaire is similar to the one described in ITU-T Rec. P.851 (2003) and is further discussed in an accompanying paper (Möller, 2004).

Because the final system was not fully integrated at the time the tests were carried out, a Wizard-of-Oz simulation of the speech recognition module and of the device control was used. Thus, a nearly perfect recognition rate could be simulated. In contrast to the recognition performance, the flow of the dialogue and the speech output are identical to the one of the final system.

During the interaction, a number of system and interaction parameters have been logged and submitted to an expert for annotation. In this way, it is possible to identify the system characteristics which are responsible for provoking particular quality percepts, to find correlations between expert-derived parameters and test subject judgements, and perhaps to predict quality percepts on the basis of interaction parameters. The test was finished in February 2004, and the analysis of the results is still ongoing.

24 test subjects participated in the Wizard-of-Oz test (19 to 29 years, mean 23.7 years). Their overall quality ratings at the end of the test (after 3 scenarios) ranged between 1.1 and 4.2 on the continuous quality rating scale (5 corresponding to the label "excellent" and 1 to the label "bad"), with a mean of 3.3. In particular, there is a large spread in the judgments between the test subjects. A further analysis will show whether this is due to the effects of different scenarios, to particular problems encountered during the interactions, or to their expectation towards the system. Still, the mean rating above the central point of the scale (corresponding to the label "fair") makes us confident that the optimized system may be acceptable for the given purpose.

## 6. Final Discussion and Conclusions

In this paper, an outline of an assessment and evaluation methodology has been presented. It has been developed for analyzing the performance of individual components of the INSPIRE smart home system, and of their contribution to overall system quality and usability. The methodology has been designed with a focus on the different environments the system will be used in, and the influence these environments carry on the performance of the components. In this way, performance values and quality judgments have been obtained which are representative for the later system.

A complete analysis of the interrelationship between system component performance and quality judgments obtained from the users is however still pending. Previous investigations show that the correlation between performance metrics and quality judgements are often very low (Möller, 2003), and that models for predicting quality from performance metrics (e.g. the PARADISE framework, see Walker et al., 1997) have only limited prediction accuracy. Thus, subjective evaluation methods like the ones presented here will persist for measuring the quality of spoken dialogue systems.

**Acknowledgements**

The present work was enabled by the European IST project INSPIRE, see http://www.inspire-project.org.